
\input harvmac
 %
\catcode`@=11   
 %

 %

\def\intem#1{\par\leavevmode%
              \llap{\hbox to\parindent{{#1}\hfill}}\ignorespaces}
 %
\def\niche#1#2{\setbox0=\hbox{#2}\hangindent\wd0\hangafter-#1%
                \par\noindent\hskip-\wd0\hbox to\wd0{\box0}}

 %

 %
\def\,{\hskip1.5pt}
 %
 %
\let\a=\alpha

\let\c=\chi
       \let\vd=\partial             
     
\let\f=\phi                       \let\F=\Phi

\let\l=\lambda

                         \let\P=\Pi
       \let\vq=\vartheta            \let\Q=\Theta

 %

\def\con{\ifmmode \hbox{\bf*} \else{\bf*}\fi}   
\def\bo{{\raise.15ex\hbox{\large$\Box\kern-.39em$}}}

\def\dual{\relax\leavevmode\lower.4ex\hbox{\titlerms*}}
\def\define{\buildrel\rm def\over =}

\let\8=\otimes
 %

 %

\let\2=\underline
\let\ha=\widehat
\let\Tw=\widetilde

\font\eightrm=cmr8
\def\6(#1){\relax\leavevmode\hbox{\eightrm(}#1\hbox{\eightrm)}}
\def\BM#1{\relax\leavevmode\setbox0=\hbox{$#1$}
           \kern-.025em\copy0\kern-\wd0
            \kern.05em\copy0\kern-\wd0
             \kern-.025em\raise.0433em\box0}

\def\0#1{\relax\ifmmode\mathaccent"7017{#1}     
                \else\accent23#1\relax\fi}      
\def\7#1#2{{\mathop{\null#2}\limits^{#1}}}      
\def\5#1#2{{\mathop{\null#2}\limits_{#1}}}      

\newbox\t@b@x
\def\rightarrowfill{$\m@th \mathord- \mkern-6mu
     \cleaders\hbox{$\mkern-2mu \mathord- \mkern-2mu$}\hfill
      \mkern-6mu \mathord\rightarrow$}
\def\tooo#1{\setbox\t@b@x=\hbox{\footnotesize$#1$}%
             \mathrel{\mathop{\hbox to\wd\t@b@x{\rightarrowfill}}%
              \limits^{#1}}\,}
\def\leftarrowfill{$\m@th \mathord\leftarrow \mkern-6mu
     \cleaders\hbox{$\mkern-2mu \mathord- \mkern-2mu$}\hfill
      \mkern-6mu \mathord-$}
\def\froo#1{\setbox\t@b@x=\hbox{\footnotesize$#1$}%
             \mathrel{\mathop{\hbox to\wd\t@b@x{\leftarrowfill}}%
              \limits^{#1}}\,}
\def\frac#1#2{{#1\over#2}}
\def\frc#1#2{\relax\ifmmode{\textstyle\frac{#1}{#2}}
                    \else$\frac{#1}{#2}$\fi}        
\def\inv#1{\frc{1}{#1}}                             
 %
\newskip\humongous \humongous=0pt plus 1000pt minus 1000pt
\def\caja{\mathsurround=0pt}
\newif\ifdtup
\def\panorama{\global\dtuptrue \openup2\jot \caja
     \everycr{\noalign{\ifdtup \global\dtupfalse
      \vskip-\lineskiplimit \vskip\normallineskiplimit
      \else \penalty\interdisplaylinepenalty \fi}}}

\def\eqalignno#1{\panorama \tabskip=\humongous
     \halign to\displaywidth{\hfil$\displaystyle{##}$
      \tabskip=0pt&$\displaystyle{{}##}$\hfil
       \tabskip=\humongous&\llap{$##$}\tabskip=0pt\crcr#1\crcr}}
\def\eqalignnotwo#1{\panorama \tabskip=\humongous
     \halign to\displaywidth{\hfil$\displaystyle{##}$
      \tabskip=0pt&$\displaystyle{{}##}$
       \tabskip=0pt&$\displaystyle{{}##}$\hfil
        \tabskip=\humongous&\llap{$##$}\tabskip=0pt\crcr#1\crcr}}
 %

 %
\def\inbar{\vrule height1.5ex width.4pt depth0pt}
\def\sinbar{\vrule height1ex width.35pt depth0pt}
\def\ssinbar{\vrule height.7ex width.3pt depth0pt}
\font\cmss=cmss10
\font\cmsss=cmss10 at 7pt
\def\ZZ{\relax\leavevmode
               \ifmmode\mathchoice
                      {\hbox{\cmss Z\kern-.4em Z}}
                      {\hbox{\cmss Z\kern-.4em Z}}
                      {\lower.9pt\hbox{\cmsss Z\kern-.36em Z}}
                      {\lower1.2pt\hbox{\cmsss Z\kern-.36em Z}}
               \else{\cmss Z\kern-.4em Z}\fi}
\def\Ik{\relax{\rm I\kern-.18em k}}
\def\IC{\relax\leavevmode
               \ifmmode\mathchoice
                      {\hbox{\kern.33em\inbar\kern-.3em{\rm C}}}
                      {\hbox{\kern.33em\inbar\kern-.3em{\rm C}}}
                      {\hbox{\kern.28em\sinbar\kern-.25em{\sevenrm C}}}
                      {\hbox{\kern.25em\ssinbar\kern-.22em{\fiverm C}}}
               \else{\hbox{\kern.3em\inbar\kern-.3em{\rm C}}}\fi}
\def\IP{\relax{\rm I\kern-.18em P}}
\def\IR{\relax{\rm I\kern-.18em R}}
\def\Ione{\relax{\rm 1\kern-3pt l}}

 %
\font\bigrm=cmr10 scaled \magstephalf
\def\Tr#1{\hbox{{\bigrm Tr}\kern-1.05em
                            \lower2.1ex\hbox{$\scriptstyle#1$}}\,}

\def\CP#1{\hbox{$\hbox{\IC \IP}^{#1}$}}

\def\muthstrut{\vphantom1}
\def\mutrix#1{\null\,\vcenter{\normalbaselines\m@th
        \ialign{\hfil$##$\hfil&&~\hfil$##$\hfill\crcr
            \muthstrut\crcr\noalign{\kern-\baselineskip}
            #1\crcr\muthstrut\crcr\noalign{\kern-\baselineskip}}}\,}

\def\EU{\relax\ifmmode \c_{{}_E} \else$\c_{{}_E}$\fi}
\def\TM{\relax\ifmmode {\cal T_M} \else$\cal T_M$\fi}
\def\TW{\relax\ifmmode {\cal T_W} \else$\cal T_W$\fi}
\def\CM{\relax\ifmmode {\cal T\rlap{\bf*}\!\!_M}
               \else$\cal T\rlap{\bf*}\!\!_M$\fi}
\def\hm#1#2{\relax\ifmmode H^{#1}(\ca{M},{#2})
                   \else$H^{#1}(\ca{M},{#2})$\fi}
\def\CP#1{\relax\ifmmode\IP^{#1}\else\IP$^{#1}$\fi}
\def\cP#1{\relax\ifmmode\IC{\rm P}^{#1}\else$\IC{\rm P}^{#1}$\fi}
\def\sll#1{\rlap{\,\raise1pt\hbox{/}}{#1}}
%

%

 %
\def\ie{\hbox{i.e.}\ }
\def\CFT{conformal field theory}
\def\CFTs{conformal field theories}
\def\CY{Calabi-Yau}
\def\?{d\kern-.3em\raise.64ex\hbox{-}}   
\def\9{\raise.43ex\hbox{-}\kern-.37em D} 
\def\3{\ifmmode\ldots\else$\ldots$\fi}
\def\ping{\nobreak\par\centerline{---$\circ$---}\goodbreak\bigskip}

 %

%

\def\NP#1{{\it Nucl.\,Phys.\,}{\bf#1\,}}
\def\PL#1{{\it Phys.\,Lett.\,}{\bf#1\,}}

\def\Pre#1{{\it #1\ University report}}
\def\pre#1{{\it University of #1 report}}

\def\MPL#1{{\it Mod.\,Phys.\,Lett.\,}{\bf#1\,}}

\def\CMP#1{{\it Commun.\,Math.\,Phys.\,}{\bf#1\,}}
\def\CQG#1{{\it Class.\,Quant.\,Grav.\,}{\bf#1\,}}
\def\IJMP#1{{\it Int.\,J.\,Mod.\,Phys.\,}{\bf#1\,}}

 %
\catcode`@=12                   
\def\cp#1#2{\hbox{$\IP_{#1}^{#2}$}}

 \def\Afour{\hsize=16.5truecm\vsize=24.7truecm}
 \Afour

\Title{\vbox{\baselineskip12pt\hbox{CERN-TH-6341/91}
                               \hbox{HUTMP-91/B319}
                                \hbox{UTTG-32-91}}}
      {\vbox{\centerline{A Generalized Construction
                            of Mirror Manifolds}}}

\centerline{Per Berglund}                            \vskip0mm
 \centerline{CERN, Theory Division}                  \vskip-1mm
 \centerline{CH-1211 Geneva 23, Switzerland}         \vskip0mm
 \centerline{and}                                    \vskip0mm
 \centerline{Theory Group, Department of Physics}    \vskip-1mm
\centerline{University of Texas, Austin, TX 78712}   \vskip-1mm
 \centerline{berglun\,@\,cernvm.bitnet}              \vskip0mm
\vskip .2in
 \centerline{Tristan H\"ubsch\footnote{$^{\spadesuit}$}
      {On leave from the Institute ``Rudjer Bo\v skovi\'c'',
       Zagreb, Croatia.}}                             \vskip0mm
 \centerline{Departments of Mathematics and Physics} \vskip-1mm
 \centerline{Harvard University, Cambridge, MA~02138}\vskip-1mm
 \centerline{hubsch\,@\,zariski.harvard.edu, @\,huma1.bitnet}
\vfill

\centerline{ABSTRACT}\vskip2mm
\vbox{\narrower\narrower\baselineskip=12pt
We generalize the known method for explicit construction of mirror
pairs of $(2,2)$-super\CFTs, using the formalism of Landau-Ginzburg
orbifolds.  Geometrically, these theories are realized as Calabi-Yau
hypersurfaces in weighted projective spaces. This generalization
makes it possible to construct the mirror partners of many manifolds
for which the mirror was not previously known.}

\Date{
\vbox{\line{CERN-TH-6341/91\hfill}
      \line{ 1/\number\yearltd \hfill}}}

\noblackbox

\nref\rDixon{For a review and references, see L.~Dixon: in {\it
     Superstrings, Unified Theories and Cosmology 1987},
     eds.~G.~Furlan et al.\ (World Scientific, Singapore, 1988)
     \,p.~67--127.}
\nref\rCHSW{P.~Candelas, G.~Horowitz, A.~Strominger and E.~Witten:
      \NP{B258}~(1985)~46.}
\nref\rGepner{D.~Gepner: \PL{199B}(1987)380, ``String theory on \CY\
     manifolds: the three generations case'', \Pre{Princeton} (December
     1987, unpublished).}
\nref\rGVW{B.R.~Greene, C.~Vafa and N.P.~Warner:
      \NP{B324}~(1989)~371\semi
       J.I.~Latorre and C.A.~L\"utken: \PL{222B}~(1989)~55\semi
       S.J.~Gates and T.~H\"ubsch: \PL{226}~(1989)~100,
      \NP{B343}~(1990)~741\semi
       B.R.~Greene: \CMP{130}~(1990)~335.}
\nref\rCGP{S.~Cecotti, L.~Girardello and A.~Pasquinucci:
      \NP{B328}~(1989)~701, \IJMP{A6}~(1991)~2427\semi
       S.~Cecotti: \IJMP{A6}~(1991)~1749, \NP{B355}~(1991)~755.}
\nref\rGP{B.R.~Greene and M.R.~Plesser: \NP{B338}~(1990)~15.}
\nref\rGQ{D.~Gepner and Z.~Qiu: \NP{B285}~(1987)~423.}
\nref\rGII{D.~Gepner: \NP{B296}~(1988)~757.}
\nref\rLS{M.~Lynker and R.~Schimmrigk: \PL{249B}~(1990)~237\semi
R.~Schimmrigk: {\it Mirror Symmetry in String Theory and Fractional
Transformations}, to appear in the proceedings of the PASCOS-91 Symposium,
Boston 1991}
\nref\rBriRon{B.R.~Greene and M.R.~Plesser: Cornell and Yale
       University preprints CLNS 91-1109, YCTP-P32-91.}
\nref\rBGH{P.~Berglund, B.~Greene and T.~H\"ubsch: ``Classical
     {\it vs}.\ Quantum Geometry of Compactification'',
      \pre{Texas} UTTG-21-91 (1991).}
\nref\rGriHa{P.~Griffiths and J.~Harris:
      {\it Principles of Algebraic Geometry}
       (John Wiley, New York, 1978).}
\nref\rLeVaWa{W.~Lerche, C.~Vafa and N.~Warner:
      \NP{B324}~(1989)~427.}
\nref\rVafa{C.~Vafa: \MPL{A4}~(1989)~1615.}
\nref\rArnold{V.I.~Arnold, S.M.~Gusein-Zade and A.N.~Varchenko: {\it
     Singularities of Differentiable Maps, Vol.~I}~ (Birkh\"auser,
     Boston, 1985).}
\nref\rMPR{P.~Candelas, M.~Lynker and R.~Schimmrigk:
      \NP{B341}~(1990)~383.}
\nref\rALR{P.S.~Aspinwall, C.A.~L\"utken and G.G.~Ross:
      \PL{241B}~(1990)~373.}
\nref\rROLF{R.~Schimmrigk: \PL{193B}~(1987)~175.}
\nref\rTH{T.~H\"ubsch: \MPL{A6}~(1991)~207, \CQG{8}~(1991)~L31.}
\nref\rRoll{P.~Candelas, P.S.~Green and T.~H\"ubsch:
      \NP{B330}~(1990)~49.}
\nref\rMMMM{P.~Berglund and T.~H\"ubsch: in preparation.}
\nref\rIV{K.~Intrilligator and C.~Vafa: \NP{B339}~(1990)~95.}
\nref\rROANII{S.-S.~Roan: {\it Int.\,J.\,Math.}\,{\bf2}~(1991)~439.}
\nref\rVS{C.~Vafa: \MPL{A4}~(1989)~1169.}
\nref\rROAN{S.-S.~Roan: {\it Int.\,J.\,Math.}\,{\bf1}~(1990)~211.}

\newsec{Introduction}
String vacua which lead to $N=1$ space-time supersymmetry can be
described by $(2,0)$-super\CFTs~\rDixon.  Another approach is to
consider Calabi-Yau manifolds as the classical background in which
the string is propagating \rCHSW. Although at first sight very
different, it is by now well-known that a large class of Calabi-Yau
spaces can be described in terms of $(2,2)$-super\CFTs\ (see for
example Refs.~\refs{\rGepner-\rCGP}).  In fact, it was
conjectured that the Calabi-Yau spaces come in pairs, where for two
spaces in such a pair the role of $(2,1)$-forms and $(1,1)$-forms
respectively are interchanged\footnote{$^1$} {We adopt the {\it
convention}~\rDixon\ where $(2,1)$-forms are equivalent to $U(1)$
charge-$(1,1)$ states in the \CFT\ language while $(1,1)$-forms are
analogs of charge-$(-1,1)$ states.}.  Two such theories are said to
form a mirror pair.  Although the two respective underlying \CFTs\
are isomorphic and differ only in the relative sign of the $U(1)$
currents in the $(2,2)$-super\CFT, it is far from straightforward to
explicitly construct manifolds which exhibit the above mirror
symmetry.

The first construction was given by Greene and Plesser~\rGP, who
considered the $3^5$ theory, \ie a tensor product of five $A_4$
superconformal minimal models.  The idea is to use the fact that
\rGQ\
$$
A_{k+1}/\ZZ_{k+2}~~ \cong ~~A_{k+1}~,
$$
the effect of the modding being a change of the relative sign of the
left- and right-moving $U(1)$ charge.  This procedure can be extended
to tensor products of minimal models \refs{\rGP, \rGII}. In particular,
$$
A_4^5/\ZZ_5~~ \cong ~~A_4^5~.
$$
Using the fact that a quintic hypersurface in $\CP{4}$, which we
denote by $\CP{4}[5]$, can be thought of as the conformal field
theory corresponding to $A_4^5/\ZZ_5$, where the quotient is the
GSO-type projection, one finds that $\CP{4}[5]/\ZZ_5^3$ is the mirror
to $\CP{4}[5]$.  The class of theories which can be described by a
polynomial (superpotential) of the Fermat type is however quite
small. A larger class is formed by those polynomials which can be
related to a Fermat type by a non-linear change of coordinates with a
constant Jacobian \refs{\rLS, \rBriRon}. Hence the above construction
for Fermat polynomials can also be used in this case to obtain the
mirror manifold in a straightforward manner.

In this paper we describe a construction of the mirror theory
from a $(2,2)$-super\CFT\ defined in terms of a
Landau-Ginzburg field theory. Unlike in the constructions outlined
above, we will not restrict (or even relate) the superpotential to
the Fermat type polynomial, which corresponds to
a tensor product of $A_k$-type minimal models. Instead, we consider
the rather more general class of non-singular polynomials
with the number of monomials the same as the number of coordinates.
In order to find the mirror, we need to consider quotients of another
theory whose defining polynomial is the transpose, in a sense that
will be made precise, of the original one.

We will make use of the recently established
ties~\refs{\rGVW,\rCGP,\rBGH} between the geometric point of view and
the Landau-Ginzburg orbifold approach.  Corresponding to a
hypersurface $\cal M$ in a weighted projective space
$\cp{(l_1,\ldots,l_5)}{4}$ defined by $P(x_i)=0$, the
$(2,2)$-super\CFT\ is determined by the superpotential $P(\F_i)$.
The fact that $P(\F_i)$ is the same polynomial as $P(x_i)$ leads to
some important identifications. In particular, the function ring of
the variety $\cal M$~\rGriHa\ and the chiral ring of the
corresponding Landau-Ginzburg model~\rLeVaWa\ are identical.
Moreover, the ring structure of the full $(p,q)$-cohomology on $\cal
M$ can be identified {\it in complete detail\,} with the full ring of
marginal operators of the Landau-Ginzburg {\it orbifold}---including
untwisted and twisted, $(c,c)$- and $(a,c)$-sectors~\rBGH.  Also, the
scaling symmetry and the associated GSO-type projection correspond to
the projectivity of the ambient space $\cal M$.  This allows us to
freely toggle between \CY\ hypersurfaces and the corresponding
Landau-Ginzburg orbifolds, and we make no notational distinction
between them.

The paper is organized as follows.  We first give the general
construction in Section~2.  In Section~3, we work out an explicit
example and present the general arguments and explicit computations
in verification of the mirror pairing of $\cal W$ with $\cal M$.
Section~4 contains our closing remarks, and some technical details are
left for the Appendices.

\newsec{The Construction}
Consider a smooth hypersurface ${\cal M}$ in a weighted projective
space $\cp{(l_1,\ldots,l_5)}{4}$ of dimension four.  The
generalization to other dimensions is
straightforward\footnote{$^2$}{$n>4$ is relevant when considering
theories with more than five fields, such as complete intersection
\CY\ manifolds and tensor products of models from the ADE series.}.
For ${\cal M}$ to be \CY, it must be defined as the zero-set of a
polynomial of degree $d= \sum_{i=1}^5 l_i$:
\eqn\eP{P(x_1, \ldots ,x_5)~ = ~0~,\qquad
         (x_1, \ldots ,x_5) \in \cp{(l_1,\ldots,l_5)}{4}~,}
and $x_i$ has scaling weight $l_i$. The corresponding statement for a
$(2,2)$-super\CFT\ leads to a theory with central charge $c=9$.  Let
us also define $Q_{\cal M}=\ZZ_d$ to be the scaling symmetry $Q_{\cal
M}$ (requiring $\l^d=1$),
\eqn\eSS{ x_i \mapsto \l^{l_i} x_i~~,\qquad P(x_i) \mapsto \l^d P(x_i)~,}
associated with $P$.  The associated Landau-Ginzburg orbifold is
obtained from the Landau-Ginzburg field theory with superpotential
$P$, by implementing the $\ZZ_d$ GSO-type projection. This
Landau-Ginzburg orbifold will also be denoted by ${\cal M}$ and we
note that this $\ZZ_d$ becomes the so--called `quantum symmetry' of
$\cal M$~\rVafa.  The group of all phase symmetries of $P$, excluding
$Q_{\cal M}$, is called the `geometric symmetry'\footnote{$^3$} {The
{\it full\,} geometric symmetry group will of course also contain
permutation symmetries, but we include these separately as usual.} of
${\cal M}$ and is denoted $G_{\cal M}$.

For the general analysis of Refs.~\refs{\rGVW,\rCGP,\rBGH} to apply,
we must require the system of gradients $\vd_i P$ to vanish only at
the origin $x_i=0$.  In this note, we also restrict $P(x_i)$ to be a
sum of only as many monomials as there are coordinates (five in the
present case), which is clearly the minimal choice.  Under this
minimality condition it is straightforward to extend the analysis of
Chapter~13 in Ref.~\rArnold\ and we list all 16 minimal non-singular
polynomials in Table~1; they are contained in the list obtained
previously in Ref.~\rMPR.  Of course, by allowing more than the
minimal number of monomials, more general superpotentials are obtained
and a similar study of such a larger class is under way; we hope to
report on these results in a detailed study.

Given a model $\cal M$ with one of the
superpotentials from Table~1, we now want to find the mirror model
$\cal W$.  The idea is to construct another model, ${\cal W}$,
such that the roles of the quantum symmetry and the
geometrical symmetry are interchanged~\rBriRon, that is,
\eqna\eQS
$$\eqalignno{ Q_{\cal M}~~ &\cong ~~G_{\cal W}~, & \eQS a \cr
              G_{\cal M}~~ &\cong ~~Q_{\cal W}~. & \eQS b \cr}$$
Note that ${\cal W}$ will in general be a quotient of a manifold
with the fixed points blown up.\ping

Let us now study the different cases at hand.  For the first
(Fermat-type) polynomial in Table~1, ${\cal W}$ is obtained by
dividing ${\cal M}$ by the action of $\P_{\cal M}$, the group of
phase symmetries which leave the $(3,0)$-form $\Omega$ invariant.
This is the technique used by Greene and Plesser as mentioned
previously \refs{\rGP,\rALR}.

We turn therefore to those polynomials for which dividing by the action
of $\P_{\cal M}$ does not yield the mirror.  To demonstrate the
procedure, let us describe it in detail for one of the polynomials in
Table~1.  The other cases will then follow easily (see also
Section~3 and Tables~1 and~2).

 From Table~1, we take for example
\eqn\eP{ P~~ = ~~x_1^{a_1} x_2 + x_2^{a_2} x_3 + x_3^{a_3}x_4
               + x_4^{a_4} x_5 + x_5^{a_5}.}
To this superpotential, we may associate the matrix of exponents
\eqn\ePmat{ P~ \simeq ~\left[
                        \mutrix{ a_1 &  0  &  0  &  0  &  0  \cr
                                  1  & a_2 &  0  &  0  &  0  \cr
                                  0  &  1  & a_3 &  0  &  0  \cr
                                  0  &  0  &  1  & a_4 &  0  \cr
                                  0  &  0  &  0  &  1  & a_5 \cr}
                       \right]~, }
whose columns are the degree vectors of the respective monomials of
$P$. It is convenient at this point to note that $P$ has a $\ZZ_{a_1
\cdots\,a_5}$ phase symmetry. To see this, let the charge of $x_5$ be
$\vq_5=1/a_5$. For the monomial $x_4^{a_4} x_5$ to transform with an
integral (not necessarily unity!) charge, we may choose $\vq_4={-1\over
a_4 a_5}$. Thereafter, $\vq_3={+1\over a_3 a_4 a_5}$, $\vq_2={-1\over
a_2 a_3 a_4 a_5}$, $\vq_1={+1\over a_1 a_2 a_3 a_4 a_5}$ and we have
a manifest $\ZZ_{a_1 \cdots\,a_5}$ action.

The new polynomial $\ha{P}$ is then defined to correspond to the
transposed matrix and so is
\eqn\ePhat{
    \ha{P}~~ = ~~x_5^{a_5} x_4 + x_4^{a_4} x_3
      + x_3^{a_3} x_2 + x_2^{a_2} x_1 + x_1^{a_1}~~,\qquad
    \ha{P}~ \simeq ~\left[
                        \mutrix{ a_1 &  1  &  0  &  0  &  0  \cr
                                  0  & a_2 &  1  &  0  &  0  \cr
                                  0  &  0  & a_3 &  1  &  0  \cr
                                  0  &  0  &  0  & a_4 &  1  \cr
                                  0  &  0  &  0  &  0  & a_5 \cr}
                       \right]~.}
We say that $\ha{P}$ is the transpose\footnote{$^4$} {It appears that
this transposition is related to the ($\ZZ_2$) reversal in the
relative sign between the $U(1)$-currents in the underlying
$(2,2)$-super\CFT.} of $P$.  Of course, transposing again gives back
$P$.  Apart from this transposition, $\ha{P}$ still belongs to the
same class of polynomials as $P$ (see Table~1). Most crucially, the
total phase symmetry of $\ha{P}$ is again $\ZZ_{a_1 \cdots\,a_5}$,
\ie\ $Q_{\cal M} \times G_{\cal M}$.

The zero locus of the transposed polynomial $\ha{P}$ defines a
hypersurface $\ha{\cal W}
=\cp{(\hat{l}_1,\ldots,\,\hat{l}_5)}{4}[\,\ha{d}\,]$, of which the
degree $\ha{d}$ and the weights $\ha{l_i}$ are determined by $\ha{P}$
using \eSS.  In general, Eqs.~\eQS{} will not be fulfilled because
$Q_{\ha{\cal W}} \subset Q_{\cal W}$, but is typically smaller than
$Q_{\cal W}$, see Section 3 and Table 2 for examples.  To enlarge
$Q_{\ha{\cal W}}$, we must divide $\ha{\cal W}$
by a suitable group of phase symmetries $H$ such that
\eqna\eQSM
$$
\eqalignnotwo{
  Q_{\cal W}~~ &\define ~~Q_{\ha{\cal W}} \times H~~
                          &\cong ~~G_{\cal M}~,      & \eQSM a\cr
  G_{\cal W}~~ &\define ~~ Q_{\ha{\cal W}}  / H~~
                          &\cong ~~Q_{\cal M}~.      & \eQSM b\cr}
$$
We then assert that ${\cal W} = {\widehat {\cal W}}/H$ is the mirror
of ${\cal M}$ and also check, case by case, that a suitable $H$ does
indeed exist. In Table~1, we list the transpose $\ha{P}$ for each of
the defining polynomials (superpotentials) $P$. We are not aware of a
definition of $H$ in closed form and suitable for display with the
general classes as in Table~1; see Table~2 for some examples.

We note in passing that all Fermat-type polynomials are
self-transposed.  From the point of view of our construction,
this is precisely the reason why the method of Greene and Plesser works
for the Fermat polynomials.
More generally, if a polynomial $P$ is self-transposed, whether
of the Fermat type or not\footnote{$^5$}{There exist special
self-transposed examples such as Eq.~\eP, with $l_i=l_{5-i}$
and so $a_i=a_{5-i}$. Most often however, transposition associates
weighted hypersurfaces of very different weights.}, we find that the
mirror of ${\cal M}$ is a quotient of $P$ such that Eqs.~\eQS{}
are fulfilled. This accounts for the original class of
mirror pairs~\refs{\rGP,\rALR}. \ping

With regard to the ADE~classification of the $N=2$ superconformal
minimal models, it is worth pointing out the following interesting fact.
It is easy to check that all models except $D_k$ ($k>3$)
are invariant under transposition. In particular, the $E_7$ polynomial
($x^3+x\,y^3$) is also self-transposed. Since it occurs in
the $1{\cdot}16^3$ model, this explains why the mirror of the
Schimmrigk manifold \rROLF, $\cal M$, is just ${\cal M}/\P_{\cal M}$.

For $D_k$ we have the following defining polynomial:
\eqn\eDK{ P(D_k)~~ = ~~x_1^{k-1} + x_1 x_2^2~, }
the transpose of which is
\eqn\eDKF{ \ha{P}(D_k)~~ = ~~x_2^2 + x_2 x_1^{k-1}~. }
For $k>3$ we see that $P\ne\ha{P}$, and so the mirror cannot in
general be found by the standard technique. That is, explicit
calculation of the $(c,c)$- and $(a,c)$-rings of relevant operators
shows that unless $k$ is even no quotient of $P(D_k)$ can be
identified with the mirror of $P(D_k)$. Instead, there is a
$\ha{P}(D_k)/\ZZ_{2(k-1)}$ Landau-Ginzburg orbifold which may be
identified with the mirror of $P(D_k)$ (the details are presented in
Appendix~A).
\footnote{$^6$}{Note that if we think of the $D_k$ model as a $\ZZ_2$ orbifold
of $A_{2k-3}$ and a trivial quadratic piece
the mirror of $D_k$ is then
given in a straightforward manner as a $\ZZ_{2(k-1)}$ orbifold of
$A_{2k-3}+A_1$.}

Given our short discussion of the $P(D_k)$ and
$\ha{P}(D_k)/\ZZ_{2(k-1)}$ Landau-Ginzburg models and the details in
Appendix~A, it is easy to see that the basic argument for the mirror
pairing of $\cal W$ with $\cal M$ is the same as in the usual
case~\rGP. There, the mirror map followed from two facts. Firstly,
\eqn\eA{ P(A_k)/\ZZ_k~~ \approx ~~P(A_k)~, }
for a 1-variable $A$-type Landau-Ginzburg model. Second, such models
could be combined into a $c=9$ theory, where the isomorphism~\eA\
reverses the sign of the Euler characteristic.  Likewise here, we
break up the polynomials in Table~1 into irreducible $n$-variable
models ($n=2,\ldots,5$) and generalizing the above $P(D_k) \approx
\ha P(D_k)/\ZZ_{2(k-1)}$ paradigm, we find that the transposition and
quotient by a suitable $H$ provide the generalization of the
isomorphism~\eA\ for all the irreducible $n$-variable models used in
Table~1. The application of this to the ``compound'' polynomials in
Table~1---and so also the mirror pairing of $\cal W$ with $\cal
M$---is then straightforward.

Greene and Plesser recently described a somewhat larger class of
models \rBriRon. They considered mirror pairs $({\cal M},{\cal W})$,
where ${\cal W}$ is constructed by first performing a fractional but
holomorphic change of coordinates to get to a Fermat polynomial. They
then deform the latter so as to obtain ${\cal W}$, with Eqs.~\eQS{}
satisfied.

Although it is always possible to get from one of the polynomials in
Table~1 to a Fermat polynomial, by a fractional holomorphic coordinate
transformation, in general it will not be possible to also deform to
a ${\cal W}$ such that the roles of the quantum and geometric
symmetry are interchanged.  Thus, the construction described in this
paper generalizes the known method for finding the mirror manifold
and makes the $\ZZ_2$ transposition manifest.

\newsec{An Example}
To illustrate the method let us consider the following example.  The
manifold below is listed in Ref.~\rMPR\ as one without a mirror.  Let
${\cal M}$ be a degree-75 hypersurface, ${\cal M}=
\cp{(5,8,12,15,35)}{4}[75]_{+6}^{27,30}$, with $\EU={+}6$,
$b_{2,1}=27$, and $b_{1,1}=30$, defined by the zero locus of
\eqn\ePE{ P~~ = ~~x_1^{15} + x_2^5 x_5 + x_1 x_5^2
                 + x_3^5 x_4 + x_4^5~.}
Note that $Q_{\cal M} = \ZZ_{75}$ and $G_{\cal M} = \ZZ_{25}
 \times \ZZ_2$.
To $P$ we can associate the following matrix (see Eq.~\ePmat)
\eqn\ePmatex{ P~~ \simeq
   ~~\left[ \mutrix{ 15  &  0  &  0  &  0  &  1  \cr
                      0  &  5  &  0  &  0  &  0  \cr
                      0  &  0  &  5  &  0  &  0  \cr
                      0  &  0  &  1  &  5  &  0  \cr
                      0  &  1  &  0  &  0  &  2  \cr} \right]~
  = ~\left[ \mutrix{ 15  &  0  &  1  \cr
                      0  &  5  &  0  \cr
                      0  &  1  &  2  \cr}\right]~ \oplus
    ~\left[ \mutrix{  5  &  0  \cr
                      1  &  5  \cr} \right]~. }

Next, we construct
\eqn\ePHAT{ \ha{P}~~ = ~~x_1^{15} x_5 + x_5^2 x_2
                        + x_2^5 + x_4^5 x_3 + x_3^5~,}
corresponding to the transposed matrix.  Using Eq.~\eP, we find that
$\ha{P} = 0$ defines a hypersurface ${\ha{\cal
W}}~=~\cp{(1,5,5,4,10)}{4}[25]_{-120}^{69,9}$.  Note that
$Q_{\widehat {\cal W}}=\ZZ_{25}$ and $G_{\widehat {\cal W}}=\ZZ_2
\times \ZZ_{75}$.  Thus $Q_{\ha{\cal W}}$ is not isomorphic to
$G_{\cal M}$.  However, on dividing by\footnote{$^7$} {We will use
the notation $(\ZZ_k:r_1,r_2,r_3,r_4,r_5)$ for a $\ZZ_k$ symmetry
with the action\hfill\break $(x_1,x_2,x_3,x_4,x_5) \to (\a^{r_1}
x_1,\a^{r_2} x_2, \a^{r_3} x_3,\a^{r_4} x_4, \a^{r_5} x_5)$, where
$\a^k = 1$.}
\eqn\eOONE{ H~~ \define ~~(\ZZ_2:1,0,0,0,1)~,}
we find that ${\cal W} = {\ha{\cal W}}/H$ satisfies
Eqs.~\eQS{} and identify ${\cal W}$ as the mirror of ${\cal M}$.

Using the method described in Appendix~B, it is now straightforward to
find not only $\EU$ but also $b_{2,1}$ and $b_{1,1}$.  For the
example in Section 3, we find $\EU({\cal W}) = {-}6$ and that
$b_{2,1} = 30$ and $b_{1,1} = 27$. Hence, as expected from the
general arguments above, ${\cal W}$ satisfies the basic numerical
requirements for the mirror of ${\cal M}$. \ping

One may wonder if there exists a polynomial whose zero locus in a
weighted projective space is the manifold ${\cal W}$ as constructed
in the above example.  After all, taking the quotient by $H$ may be
enforced by the following fractional change of coordinates:
\eqn\eFHT{ (x_1 \,,\, x_2 \,,\, x_3 \,,\, x_4 \,,\, x_5)~~ =
          ~~(\tilde x_1^{1/2} ,\, \tilde x_2 \,,\, \tilde x_3 \,,\,
              \tilde x_4 \,,\, \tilde x_5 \tilde x_1^{1/2})~. }
This gives us the  polynomial
\eqn\ePTIL{\Tw{P}~~ = ~~\tilde x_1^8 \tilde x_5 +
                         \tilde x_5^2 \tilde x_1\tilde x_2 +
                          \tilde x_2^5 + \tilde x_4^5 \tilde x_3 +
                           \tilde x_3^5~.}
It is easy to check that $\Tw{P} = 0$ describes a degree-25
hypersurface in $\widetilde {\cal W}=\cp{(2,5,5,4,9)}{4}$.  However,
$\Tw{P}$ is not transverse at $p^{\sharp} =(0,0,0,0,1)$.  Hence we
cannot expect Vafa's formula for the Euler number \rVS~
(see also Eq. (B.1)) to give us the correct spectrum as it only
holds for smooth manifolds, that is non-degenerate Landau-Ginzburg
orbifolds. Note however that $p^{\sharp}$ is in the fixed point set
of $H$ and is also a local $\ZZ_9$-singularity as inherited from the
weighted projective space. Interpreting (as usual) the
Landau-Ginzburg orbifold as a blow-up at the fixed point set, the
singularity may actually be smoothed completely and the situation may
not be as bad as it seems.  In a local coordinate system with
$\tilde{x}_5=1$ and $u=(\tilde{x}_4^5+\tilde{x}_3^4)$ at
$p^{\sharp}$, we have
\eqn\elocWtil{ \Tw{P}\big|_{p^{\sharp}}~~ =
              ~~\tilde x_1^8 + \tilde x_1\tilde x_2 + \tilde x_2^5
                  + u\, \tilde x_3~,\qquad
               \det[\vd^2\Tw{P}]\big|_{p^{\sharp}}~~ = ~~1~. }
The singularity is therefore also a node, the Landau-Ginzburg
orbifold is degenerate and we are unable to compute the numerical
characteristics reliably~\rTH.

In spite of the technical problems in computing the correct spectrum
for a non-transverse polynomial, we believe that $\Tw{\cal W}$ indeed
is the mirror to ${\cal M}$. We base this conjecture on the fact that
an appropriate resolution of the singular $\Tw{\cal W}$ may be
identified with the quotient $\ha{\cal W}/H$. Note, however, that
 $\Tw{\cal W}$ and ${\cal W}$ will correspond in
general to two different
points in the moduli space. This would lead to a special class of
\CY\ conifolds~\rRoll, which are non-transverse precisely at the fixed
points of the scaling symmetry $Q_{\widetilde {\cal W}}$. It would be
interesting to list all non-transverse polynomials of the above type
to see if they would complete the list of weighted projective spaces
\rMPR---as far as providing every manifold there with its mirror.

In Table~2, we give several more examples of theories for which the
mirror was not previously known. As well as being of potential
phenomenological interest, we hope that they will illustrate the ease
with which the mirror model is constructed.

\newsec{Discussion and Conclusions}

The cautious reader may worry whether there is a hidden caveat to our
argument for the mirror pairing of $\cal W$ with $\cal M$.  Indeed,
the only completely unambiguous proof would have us compute the
(correctly) normalized three-point functions $\vev{\f_{(1,1)}^3}$ and
$\vev{\f_{(-1,1)}^3}$ in the underlying $(2,2)$-super\CFT. If
\eqn\ePHI{ \vev{\f_{(1,1)}^3}_{\cal M}~
          =~\vev{\f_{({-}1,1)}^3}_{\cal W}~, \qquad
           \vev{\f_{(1,-1)}^3}_{\cal M}~
          =~\vev{\f_{(1,1)}^3}_{\cal W}~,}
the proof is complete, since the three-point functions completely
determine the $(2,2)$-super\CFT.

Unfortunately, it does not seem possible at this time to complete the
above calculation, because the $(2,2)$-super\CFTs\ we find are not
tensor products of minimal models and the correct normalizations seem
to elude us.  Nevertheless, using the ring structure of the
Landau-Ginzburg orbifolds, specified through the Jacobian ideal
generated by the system of gradients of the superpotential, as well
as the quantum and geometric symmetries, it is straightforward to
show that the general structure of the Yukawa couplings, such as the
zeros, is the same.

Without delving into the calculation (which will be presented in full
detail elsewhere \rMMMM), for the example discussed earlier we find a
one-to-one relation between the $(1,1)$ states in the super\CFT\
corresponding to ${\cal M}$ and the $(-1,1)$ states in ${\cal W}$ and
vice versa.

We have described a technique which generalizes the existing methods
of obtaining the mirror manifold to a given hypersurface in a
weighted projective space.  The method also applies for complete
intersection \CY\ manifolds, for which the covering space can be
embedded in a (product of) weighted projective space(s)---as long as
the superpotential has as many monomials as there are variables.  We
conjecture that in general the mirror ${\cal W}$ to a manifold ${\cal
M}$ does not have to be described by a transverse
polynomial---although ${\cal W}$ can be expressed as a quotient of a
non-singular covering space $\ha{\cal W}$ and the non-transversality
occurs at fixed points of $Q_{\cal W}$.  The straightforward way in
which the method is applied makes us believe that one ought to be
able to mechanize the procedure in terms of a computer code.  From
the list in Ref.~\rMPR, one would then be able to obtain a very large
class of mirror pairs.  Although there may still remain manifolds
without a constructed mirror, we shall be a step closer to verifying
that every \CY\ manifold indeed has a mirror.

\vfill
{\bf Acknowledgements}:
P.B. acknowledges useful discussions with P.~Candelas , E.~Derrick,
 X.~de la Ossa and J.~Louis. P.B.\ was supported by the Foundation
Blanceflor-Boncompagni-Ludovisi n\'ee Bildt, the Fulbright Program,
 a University of Texas Fellowship, and in part by the NSF grant PHY
9009850 and the Robert~A.~Welch foundation.
T.H. was supported by the DOE grant DE-FG02-88ER-25065 and would also
like to thank the Department of Mathematics of the National Tsing-Hua
University at Hsinchu, Taiwan, for the warm hospitality during the
time when part of this research was completed.

\vfill\eject

\appendix{A}{The Mirror of $D_k$}

In this appendix, we will show that the mirror to $D_k$ is given by
$\ha P(D_k)/\ZZ_{2(k-1)}$ for $k>2$. We have that the defining
 polynomial for $D_k$ is given by
\eqn\qDk{P(D_k)=x^{k-1} + x y^2~.}
The charges are $q_x={1\over k-1}$ and $q_y={k - 2\over 2(k-1)}$.
The corresponding Landau-Ginzburg theory has the following ring
structure
$$
\eqalignno{
(c,c)~:~~ &\{\> |0\rangle^{(0)}_{(c,c)}, ~x|0\rangle^{(0)}_{(c,c)},
                  ~\ldots, ~x^{k-2}|0\rangle^{(0)}_{(c,c)},
                            ~y|0\rangle^{(0)}_{(c,c)} \>\}\strut_k \cr
(a,c)~:~~ &\{\> |0\rangle^{(0)}_{(a,c)} \>\}\strut_1}
$$
where the subscripts $k$ and $1$ indicate the number of states in the
$(c,c)$ and $(a,c)$ rings respectively.  To find the mirror theory we
consider quotients of $D_k$. Depending on whether $k$ is even or odd,
there are two situations.

For $k$ even, the scaling symmetry $j$ of $D_k$ is
$(\ZZ_{k-1}:1,{k-2\over2})$. Using the result in Ref.~\rIV\ for the
transformation of a given state under a symmetry, we find the states
which are invariant under $j$. This gives rise to the ring structure
$$
\eqalignno{
(c,c)~:~~ &\{\> |0\rangle^{(0)}_{(c,c)} \>\}\strut_1 \cr
(a,c)~:~~ &\{\> |0\rangle^{(0)}_{(a,c)}, ~|0\rangle^{(k-2)}_{(a,c)},
                ~\ldots, ~|0\rangle^{(2)}_{(a,c)},
                           ~x^{{k\over 2}-1}|0\rangle^{(1)}_{(a,c)},
                             ~y|0\rangle^{(1)}_{(a,c)} \>\}\strut_k}
$$
It is easy to check, using Eq.~(B.4), that the $(a,c)$ states above
have the same charges as the $(c,c)$ states in the $D_k$ theory
except that $q_L \to -q_L$.  So $D_k/\ZZ_{k-1}$ is the mirror to
$D_k$ for $k$ even.

When $k$ is odd the scaling symmetry $j$ is $(\ZZ_{2(k-1)}:2,k{-}2)$.
A calculation similar to the one above shows that, unless $k=3$,
$D_k/j$ is not the mirror; neither is any other quotient. We then
turn to the transposed polynomial for $D_k$,
\eqn\eDkH{\ha P(D_k)=y^2 + y x^{k-1}~,}
which has charges $q_1={1\over 2(k-1)}$ and $q_2={1\over 2}$. Note that
for all $k$ (even and odd), the scaling symmetry is $\ha\jmath =
(\ZZ_{2(k-1)}:1, k{-}1)$. It is then straightforward to obtain the
ring structure for the LG-orbifold $\ha P(D_k)/\ha\jmath$~:
$$
\eqalignno{
(c,c)~:~~ &\{\> |0\rangle^{(0)}_{(c,c)} \>\}\strut_1 \cr
(a,c)~:~~ &\{\> |0\rangle^{(0)}_{(a,c)},
                ~|0\rangle^{(2(k-1)-2)}_{(a,c)}, ~\ldots,
                 ~|0\rangle^{(2)}_{(a,c)},
                  ~y|0\rangle^{(1)}_{(a,c)} \>\}\strut_k~.}
$$
Note that for $\ell=2p+1$, the states
$x^{k-2}|0\rangle^{(1)}_{(a,c)}$ are not projected out, but $\del_x
P(D_k)|_{y=0}=x^{k-2}$ and the state is in the ideal. (We restrict to
$y=0$ because $\Q_y(2p+1) \not\in \ZZ$.)  Using Eq.~\eQS{}, one can
check that the charges are the same as for the $(c,c)$ states in the
$D_k$ model, with a change of sign for $q_L$. Thus,
$\ha{P}(D_k)/\ha\jmath$ is the mirror of $D_k$ for any $k>2$.

\vfill
\eject
\appendix{B}{Computation of $\EU$, $b_{2,1}$ and $b_{1,1}$}
We next turn to compute the $\EU$, $b_{2,1}$ and $b_{1,1}$ for
${\cal W}$ (see also Ref.~\rROANII). To this end, we need the
expression for the Euler number for a weighted projective
hypersurface~\refs{\rVS,\rROAN}~:
\eqn\ePI{ \EU~~ = ~~{1\over d}
             \sum_{r,\ell=0}^{d-1} (-1)^{(D-N) (r + \ell + r\ell)}
              \prod_{\ell q_i, rq_i \in \ZZ} \left(1-{1\over q_i}
              \right)~,}
where $D=3$ is the dimension of the \CY\ space, $N=5$ is the number
of homogeneous coordinates, and $q_i = \ell_i/d$.
We now want to generalize Eq.~\ePI\ so that it can also be valid
for a quotient. First, rewrite it as
\eqn\ePIP{ \EU~~ = ~~\sum_{r=0}^{d-1} S_r~,}
where
\eqn\eSTATES{ S_r~~ = ~~{1\over d}
            \sum_{\ell=0}^{d-1} (-1)^{(D-N) (r + \ell + r\ell)}
             \prod_{\Q_i(\ell), \Q_i(r) \in \ZZ}
               \left(1-{1\over q_i}\right)~.}
Note that $r$ and $\ell$ run over all twisted sectors, including the
sectors due to dividing by the discrete symmetry;  $\Q_i(\ell)$ is
the $i^{th}$ twist charge from the $\ell^{th}$ twisted sector and
similarly for $\Q_i(r)$. This is a generalization of $\ell q_i$ and
$rq_i$ in Eq.~\ePI~\refs{\rIV,\rBGH}.  The point of writing $\EU$ in
the form \ePIP\ is that it resembles the usual expression for the
Euler number.  In fact, the $S_r$ will determine the Hodge numbers.

To extract $b_{2,1}$ and $b_{1,1}$, we need to know which sectors
contribute to $(2,1)$- and $(1,1)$-forms, respectively. The simplest
way to see this is by looking at the \CFT.  Since $\EU$ is given by
Eq.~\ePI, it is enough to consider $b_{2,1}$.  Recall that we
associate charge-$(1,1)$ states to $(2,1)$-forms.  The question is
then which sectors contain $(1,1)$ states. To answer this, we need
the expression for the charges of the Ramond vacuum in the
$\ell^{th}$ twisted sector~\rVS~:
\eqn\eQR{ {J_0\atop \bar J_0} \Big|0\Big\rangle^{(\ell)}_R~ = \left\{
   \pm\left[\sum\limits_{\Q_i(\ell) \not\in \ZZ}
      \left(\Q_i(\ell) - [\Q_i(\ell)] - {1\over2}\right)\right]~ +
     ~\left[\sum\limits_{\Q_i(\ell) \in \ZZ}
       \left(q_i - {1\over2}\right)\right] \,\right\}
                                  \Big|0\Big\rangle^{(\ell)}_R~ .}
The $|0\rangle^{(\ell)}_{(c,c)}$ vacua are obtained by spectral flow
${\cal U}_{(1/2,1/2)}$, of charges $({c\over6},{c\over6})$.  It is
then an easy exercise to find those sectors that contain marginal
states of charge $(1,1)$.  So we find that \eqna\eBN
$$
\eqalignnotwo{
 b_{2,1}~ &= ~\inv{2} b_3 - 1~~
          &= ~~{-}{1\over2} \sum_{r_+} S_{r_+}~ - ~1~, & \eBN a\cr
 b_{1,1}~ &= ~~{1\over2}\EU - b_{2,1}
          &= ~~{1\over2} \sum_{r\not=r_+} S_r~ + ~1~,  & \eBN b\cr}
$$
where $r_+$ runs over the sectors contributing charge-$(1,1)$ states.
The reason for the form of the expression for $b_{2,1}$ is that
$S_{r_+}$ also contains an equal number of $(2,2)$ states as well as
one $(0,0)$ and one $(3,3)$ state respectively from the untwisted
sector, $p = 0$. The analogy between 3-forms and integral
$(c,c)$-states is self-evident and well-known.

\vfill
\eject

\vbox{
$$\vbox{\offinterlineskip
\hrule height 1.1pt
\halign{&\vrule width 1.1pt#&\strut\quad#\hfil\quad&
\vrule#&\strut\quad#\hfil\quad&\vrule width 1.1pt#\cr
height6pt&\omit&&\omit&\cr
&\hfil$P$&&$\hfil\widehat{P}$&\cr
height6pt&\omit&&\omit&\cr
\noalign{\hrule  height 1.1pt\vskip3pt\hrule height 1.1pt}
height6pt&\omit&&\omit&\cr
&$x_1^{a_1} + x_2^{a_2} + x_3^{a_3} + x_4^{a_4} + x_5^{a_5}$&
&$x_1^{a_1} + x_2^{a_2} + x_3^{a_3} + x_4^{a_4} + x_5^{a_5}$&\cr
height6pt&\omit&&\omit&\cr
\noalign{\hrule}
height6pt&\omit&&\omit&\cr
&$x_1^{a_1} x_2 + x_2^{a_2} + x_3^{a_3} + x_4^{a_4} + x_5^{a_5}$&
&$x_2^{a_2} x_1 + x_1^{a_1} + x_3^{a_3} + x_4^{a_4} + x_5^{a_5}$&\cr
height6pt&\omit&&\omit&\cr
\noalign{\hrule}
height6pt&\omit&&\omit&\cr
&$ x_1^{a_1} x_2 + x_2^{a_2} x_3 + x_3^{a_3} + x_4^{a_4} + x_5^{a_5}$&
&$ x_3^{a_3} x_2 + x_2^{a_2} x_1 + x_1^{a_1} + x_4^{a_4} + x_5^{a_5}$&\cr
height6pt&\omit&&\omit&\cr
\noalign{\hrule}
height6pt&\omit&&\omit&\cr
&$x_1^{a_1} x_2 + x_2^{a_2} x_3 + x_3^{a_3}x_4 +
          x_4^{a_4} + x_5^{a_5}$&
&$x_4^{a_4} x_3 + x_3^{a_3} x_2 + x_2^{a_2}x_1 +
          x_1^{a_1} + x_5^{a_5}$&\cr
height6pt&\omit&&\omit&\cr
\noalign{\hrule}
height6pt&\omit&&\omit&\cr
&$x_1^{a_1} x_2 + x_2^{a_2} x_3 + x_3^{a_3}x_4 +
          x_4^{a_4} x_5 + x_5^{a_5}$&
&$x_5^{a_5} x_4 + x_4^{a_4} x_3 + x_3^{a_3}x_2 +
          x_2^{a_2} x_1 + x_1^{a_1}$&\cr
height6pt&\omit&&\omit&\cr
\noalign{\hrule}
height6pt&\omit&&\omit&\cr
&$ x_1^{a_1} x_2 + x_2^{a_2} + x_3^{a_3} x_4 + x_4^{a_4} + x_5^{a_5}$&
&$ x_2^{a_2} x_1 + x_1^{a_1} + x_4^{a_4} x_4 + x_3^{a_3} + x_5^{a_5}$&\cr
height6pt&\omit&&\omit&\cr
\noalign{\hrule}
height6pt&\omit&&\omit&\cr
&$ x_1^{a_1} x_2 + x_2^{a_2} x_3 + x_3^{a_3} +
 x_4^{a_4} x_5 + x_5^{a_5}$&
&$ x_3^{a_3} x_2 + x_2^{a_2} x_1 + x_1^{a_1} +
 x_5^{a_5} + x_5 x_4^{a_4}$&\cr
height6pt&\omit&&\omit&\cr
\noalign{\hrule}
height6pt&\omit&&\omit&\cr
&$x_1^{a_1} + x_2^{a_2} + x_3^{a_3} + x_4^{a_4} x_5 +
            x_5^{a_5} x_4$&
&$x_1^{a_1} + x_2^{a_2} + x_3^{a_3} + x_4^{a_4} x_5 +
            x_5^{a_5} x_4$&\cr
height6pt&\omit&&\omit&\cr
\noalign{\hrule}
height6pt&\omit&&\omit&\cr
&$x_1^{a_1} + x_2^{a_2} x_3  + x_3^{a_3} + x_4^{a_4} x_5 +
            x_5^{a_5} x_4$&
&$x_1^{a_1} + x_3^{a_3} x_2  + x_2^{a_2} + x_4^{a_4} x_5 +
            x_5^{a_5} x_4$&\cr
height6pt&\omit&&\omit&\cr
\noalign{\hrule}
height6pt&\omit&&\omit&\cr
&$x_1^{a_1} x_2 + x_2^{a_2} x_3  + x_3^{a_3} + x_4^{a_4} x_5 +
            x_5^{a_5} x_4$&
&$x_3^{a_3} x_2 + x_2^{a_2} x_1  + x_1^{a_1} + x_4^{a_4} x_5 +
            x_5^{a_5} x_4$&\cr
height6pt&\omit&&\omit&\cr
\noalign{\hrule}
height6pt&\omit&&\omit&\cr
&$x_1^{a_1} + x_2^{a_2} x_3  + x_3^{a_3} x_2 + x_4^{a_4} x_5 +
            x_5^{a_5} x_4$&
&$x_1^{a_1} + x_2^{a_2} x_3  + x_3^{a_3} x_2 + x_4^{a_4} x_5 +
            x_5^{a_5} x_4$&\cr
height6pt&\omit&&\omit&\cr
\noalign{\hrule}
height6pt&\omit&&\omit&\cr
&$x_1^{a_1} + x_2^{a_2} + x_3^{a_3} x_4 + x_4^{a_4} x_5 +
                   x_5^{a_5} x_3$&
&$x_1^{a_1} + x_2^{a_2} + x_5^{a_5} x_4 + x_4^{a_4} x_3 +
                   x_3^{a_3} x_5$&\cr
height6pt&\omit&&\omit&\cr
\noalign{\hrule}
height6pt&\omit&&\omit&\cr
&$x_1^{a_1} x_2 + x_2^{a_2} + x_3^{a_3} x_4 + x_4^{a_4} x_5 +
                   x_5^{a_5} x_3$&
&$x_2^{a_2} x_1 + x_1^{a_1} + x_5^{a_5} x_4 + x_4^{a_4} x_3 +
                   x_3^{a_3} x_5$&\cr
height6pt&\omit&&\omit&\cr
\noalign{\hrule}
height6pt&\omit&&\omit&\cr
&$x_1^{a_1} x_2 + x_2^{a_2} x_1 + x_3^{a_3} x_4 + x_4^{a_4} x_5 +
                   x_5^{a_5} x_3$&
&$x_1^{a_1} x_2 + x_2^{a_2} x_1 + x_5^{a_5} x_4 + x_4^{a_4} x_3 +
                   x_3^{a_3} x_5$&\cr
height6pt&\omit&&\omit&\cr
\noalign{\hrule}
height6pt&\omit&&\omit&\cr
&$x_1^{a_1} + x_2^{a_2} x_3 + x_3^{a_3} x_4 + x_4^{a_4} x_5 +
           x_5^{a_5} x_2$&
&$x_1^{a_1} + x_5^{a_5} x_4 + x_4^{a_4} x_3 + x_3^{a_3} x_2 +
           x_2^{a_2} x_5$&\cr
height6pt&\omit&&\omit&\cr
\noalign{\hrule}
height6pt&\omit&&\omit&\cr
&$x_1^{a_1} x_2 + x_2^{a_2} x_3 + x_3^{a_3} x_4 + x_4^{a_4} x_5 +
           x_5^{a_5} x_1$&
&$x_5^{a_5} x_4 + x_4^{a_4} x_3 + x_3^{a_3} x_2 + x_2^{a_2} x_1 +
           x_1^{a_1} x_5$&\cr
height6pt&\omit&&\omit&\cr}
\hrule height 1.1pt}$$
\vskip7pt
\noindent
{\bf Table 1}: $P=0$ defines a hypersurface in a weighted projective
4-space. $\ha{P}=0$ gives the covering space, $\ha{\cal W}$,
of ${\cal W}$, the mirror of ${\cal M}$.
${\cal W}$ will in general be a quotient of $\ha{\cal W}$.}

\vfill \eject

\vbox{
$$\vbox{\offinterlineskip
\hrule height 1.1pt
\halign{&\vrule width 1.1pt#&\strut\quad#\hfil\quad&
\vrule#&\strut\quad#\hfil\quad&
\vrule#&\strut\quad#\hfil\quad&\vrule width 1.1pt#\cr
height7pt&\omit&&\omit&&\omit&\cr
&\hfil${\cal M~,~\widehat W}$&&\hfil$P~,~\ha{P}$&&$\hfil{\cal W}$&\cr
height7pt&\omit&&\omit&&\omit&\cr
\noalign{\hrule height 1.1pt\vskip3pt\hrule height 1.1pt}
height7pt&\omit&&\omit&&\omit&\cr
&$\cp{(2,8,29,49,59)}{4}[147]^{51,48}_{-6}$&
&$x_2^{11} x_5 {+} x_5^2 x_3 {+} x_3^5 x_1 {+} x_1^{49} x_4 {+} x_4^3$&&&\cr
height7pt&\omit&&\omit&&\omit&\cr
&$\cp{(1,5,6,18,25)}{4}[55]^{97,28}_{-138}$&
&$x_4^3 x_1 {+} x_1^{49} x_3 {+} x_3^5 x_5 {+} x_5^2 x_2 {+} x_2^{11}$&
&$\widehat {\cal W}/(\ZZ_2:1,0,1,1,1)$&\cr
height7pt&\omit&&\omit&&\omit&\cr
\noalign{\hrule}
height7pt&\omit&&\omit&&\omit&\cr
&$\cp{(3,4,14,21,21)}{4}[63]^{35,32}_{-6}$&
&$x_2^{15} x_1 {+} x_1^{21} {+} x_3^3 x_4 {+} x_4^3 {+} x_5^3$&&&\cr
height7pt&\omit&&\omit&&\omit&\cr
&$\cp{(2,3,10,15,15)}{4}[45]^{49,22}_{-54}$&
&$x_1^{21} x_2 {+} x_2^{15} {+} x_4^3 x_3 {+} x_3^3 {+} x_5^3$&
&$\widehat {\cal W}/(\ZZ_3:1,0,1,1,0)$&\cr
height7pt&\omit&&\omit&&\omit&\cr
\noalign{\hrule}
height7pt&\omit&&\omit&&\omit&\cr
&$\cp{(3,5,8,24,35)}{4}[75]^{43,40}_{-6}$&
&$x_1^{25} {+} x_2^{15} {+} x_3^5 x_5 {+} x_4^3 x_1 {+} x_5^2 x_2$&&&\cr
height7pt&\omit&&\omit&&\omit&\cr
&$\cp{(2,3,15,25,30)}{4}[75]^{75,27}_{-96}$&
&$x_1^{25} x_4 {+} x_4^3 {+} x_2^{15} x_5 {+} x_5^2 x_3 {+} x_3^5$&
&$\widehat {\cal W}/(\ZZ_2:0,1,0,0,1)$&\cr
height7pt&\omit&&\omit&&\omit&\cr
\noalign{\hrule}
height7pt&\omit&&\omit&&\omit&\cr
&$\cp{(4,5,26,65,95)}{4}[195]^{67,70}_{+6}$&
&$x_1^{25} x_5 {+} x_5^2 x_2 {+} x_2^{39} {+} x_3^5 x_4 {+} x_4^3$&&&\cr
height7pt&\omit&&\omit&&\omit&\cr
&$\cp{(1,3,15,20,36)}{4}[75]^{145,31}_{-228}$&
&$x_2^{39} x_5 {+} x_5^2 x_1 {+} x_1^{25} {+} x_4^3 x_3 {+} x_3^5$&
&$\widehat {\cal W}/(\ZZ_2:1,0,0,0,1)$&\cr
height7pt&\omit&&\omit&&\omit&\cr
\noalign{\hrule}
height7pt&\omit&&\omit&&\omit&\cr
&$\cp{(5,6,14,45,65)}{4}[135]^{42,45}_{+6}$&
&$x_3^5 x_5 {+} x_5^2 x_1 {+} x_1^{27} {+} x_2^{15} x_4 {+} x_4^3$&&&\cr
height7pt&\omit&&\omit&&\omit&\cr
&$\cp{(1,3,9,14,18)}{4}[45]^{95,23}_{-144}$&
&$x_1^{27} x_5 {+} x_5^2 x_3 {+} x_3^5 {+} x_4^3 x_2 {+} x_2^{15}$&
&$\widehat {\cal W}/(\ZZ_2:1,0,0,0,1)$&\cr
height7pt&\omit&&\omit&&\omit&\cr
\noalign{\hrule}
height7pt&\omit&&\omit&&\omit&\cr
&$\cp{(5,8,12,15,35)}{4}[75]^{27,30}_{+6}$&
&$x_2^5 x_5 {+} x_5^2 x_1 {+} x_1^{15} {+} x_3^5 x_4 {+} x_4^5$&&&\cr
height7pt&\omit&&\omit&&\omit&\cr
&$\cp{(1,5,5,4,10)}{4}[25]^{69,9}_{-120}$&
&$x_1^{15} x_5 {+} x_5^2 x_2 {+} x_2^5 {+} x_4^5 x_3 {+} x_3^5$&
&$\widehat {\cal W}/(\ZZ_2:1,0,0,0,1)$&\cr
height7pt&\omit&&\omit&&\omit&\cr
\noalign{\hrule}
height7pt&\omit&&\omit&&\omit&\cr
&$\cp{(2,6,9,17,17)}{4}[51]^{31,34}_{+6}$&
&$x_4^3 {+} x_1^{17} x_5 {+} x_5^3 {+} x_2^7 x_3 {+} x_3^5 x_2$&&&\cr
height7pt&\omit&&\omit&&\omit&\cr
&$\cp{(3,6,9,17,16)}{4}[51]^{66,51}_{-102}$&
&$x_4^3 {+} x_5^3 x_1 {+} x_1^{17} {+} x_3^5 x_2 {+} x_2^7 x_3$&
&$\widehat {\cal W}/(\ZZ_2:0,1,1,0,0)$&\cr
height7pt&\omit&&\omit&&\omit&\cr
\noalign{\hrule}
height7pt&\omit&&\omit&&\omit&\cr
&$\cp{(4,4,11,17,19)}{4}[55]^{24,21}_{-6}$&
&$x_1^{11} x_3  {+} x_3^5 {+} x_2^9 x_5 {+} x_5^2 x_4 {+} x_4^3 x_2$&&&\cr
height7pt&\omit&&\omit&&\omit&\cr
&$\cp{(1,1,2,2,5)}{4}[11]^{109,4}_{-210}$&
&$x_3^5 x_1 {+} x_1^{11} {+} x_4^3 x_5 {+} x_5^2 x_2 {+} x_2^9 x_4$&
&$\widehat {\cal W}/(\ZZ_5:0,1,1,1,2)$&\cr
height7pt&\omit&&\omit&&\omit&\cr}
\hrule height 1.1pt}$$
\vskip7pt
\noindent
{\bf Table 2}: A list of ${\cal M}$ and their mirrors ${\cal W}$.
$\ha{\cal W}$ is defined by $\ha{P}=0$.
A GSO-type $\ZZ_d$-~projection is implicitly understood,
where $d$ is the degree of the defining polynomial.
${\cal M}$ are manifolds for which no mirror was listed in
Ref.~\rMPR.}

\vfill \eject

\listrefs

\bye